\documentclass[aic,crcready]{iosart2x}

\pubyear{0000}
\volume{0}
\firstpage{1}
\lastpage{1}

\begin{document}

\begin{frontmatter}

%\pretitle{}
\title{Developing, Evaluating and Scaling Learning Agents in Multi-Agent Environments}
\runtitle{Developing, Evaluating and Scaling Learning Agents in Multi-Agent Environments}
%\subtitle{}

% For one author:
%\author{\inits{N.}\fnms{Name1} \snm{Surname1}\ead[label=e1]{first@somewhere.com}}
%\address{Department first, \orgname{University or Company name},
%Abbreviate US states, \cny{Country}\printead[presep={\\}]{e1}}

% Two or more authors:
\begin{aug}
\author[A]{\inits{I.}\fnms{Ian} \snm{Gemp}\ead[label=e1]{imgemp@deepmind.com}}

\author[A]{\inits{D.}\fnms{Thomas} \snm{Anthony}\ead[label=e1]{twa@deepmind.com}}

\author[A]{\inits{Y.}\fnms{Yoram} \snm{Bachrach}\ead[label=e1]{yorambac@deepmind.com}}

\author[A]{\inits{A.}\fnms{Avishkar} \snm{Bhoopchand}\ead[label=e1]{avishkar@deepmind.com}}

\author[A]{\inits{J.}\fnms{Kalesha} \snm{Bullard}\ead[label=e1]{ksbullard@deepmind.com}}

\author[A]{\inits{J.}\fnms{Jerome} \snm{Connor}\ead[label=e1]{jeromeconnor@deepmind.com}}

\author[A]{\inits{V.}\fnms{Vibhavari} \snm{Dasagi}\ead[label=e1]{vibhadasagi@deepmind.com}}

\author[A]{\inits{B.}\fnms{Bart} \snm{De Vylder}\ead[label=e1]{bartdv@deepmind.com}}

\author[A]{\inits{E.}\fnms{Edgar~A.} \snm{Du{\'e}{\~n}ez-Guzm{\'a}n}\ead[label=e1]{duenez@deepmind.com}}

\author[A]{\inits{R.}\fnms{Romuald} \snm{Elie}\ead[label=e1]{relie@deepmind.com}}

\author[A]{\inits{R.}\fnms{Richard} \snm{Everett}\ead[label=e1]{reverett@deepmind.com}}

\author[A]{\inits{D.}\fnms{Daniel} \snm{Hennes}\ead[label=e1]{hennes@deepmind.com}}

\author[A]{\inits{E.}\fnms{Edward} \snm{Hughes}\ead[label=e1]{edwardhughes@deepmind.com}}

\author[A]{\inits{M.}\fnms{Mina} \snm{Khan}\ead[label=e1]{minakhan@deepmind.com}}

\author[A]{\inits{M.}\fnms{Marc} \snm{Lanctot}\ead[label=e1]{lanctot@deepmind.com}}

\author[A]{\inits{K.}\fnms{Kate} \snm{Larson}\ead[label=e1]{katelarson@deepmind.com}}

\author[A]{\inits{G.}\fnms{Guy} \snm{Lever}\ead[label=e1]{guylever@deepmind.com}}

\author[A]{\inits{S.}\fnms{Siqi} \snm{Liu}\ead[label=e1]{liusiqi@deepmind.com}}

\author[A]{\inits{L.}\fnms{Luke} \snm{Marris}\ead[label=e1]{marris@deepmind.com}}

\author[A]{\inits{K.R.}\fnms{Kevin R.} \snm{McKee}\ead[label=e1]{kevinrmckee@deepmind.com}}

\author[A]{\inits{P.}\fnms{Paul} \snm{Muller}\ead[label=e1]{pmuller@deepmind.com}}

\author[A]{\inits{J.}\fnms{Julien} \snm{P{\'e}rolat}\ead[label=e1]{perolat@deepmind.com}}

\author[A]{\inits{F.}\fnms{Florian} \snm{Strub}\ead[label=e1]{fstrub@deepmind.com}}

\author[A]{\inits{A.}\fnms{Andrea} \snm{Tacchetti}\ead[label=e1]{atacchet@deepmind.com}}

\author[A]{\inits{E.}\fnms{Eugene} \snm{Tarassov}\ead[label=e1]{etar@deepmind.com}}

\author[A]{\inits{Z.}\fnms{Zhe} \snm{Wang}\ead[label=e1]{zhewang@deepmind.com}}

\author[A]{\inits{K.}\fnms{Karl} \snm{Tuyls}\ead[label=e1]{karltuyls@deepmind.com}%
\thanks{Corresponding author. \printead{e1}.}}

\address[A]{Game Theory \& Multi-Agent Team, \orgname{DeepMind},
London, \cny{UK}\printead[presep={\\}]{e1}}
% \address[B]{Department first, \orgname{University or Company name},
% Abbreviate US states, \cny{Country}\printead[presep={\\}]{e2,e3}}
\end{aug}

\begin{abstract}
The Game Theory \& Multi-Agent team at DeepMind studies several aspects of multi-agent learning ranging from computing approximations to fundamental concepts in game theory to simulating social dilemmas in rich spatial environments and training 3-d humanoids in difficult team coordination tasks. A signature aim of our group is to use the resources and expertise made available to us at DeepMind in deep reinforcement learning to explore multi-agent systems in complex environments and use these benchmarks to advance our understanding. Here, we summarise the recent work of our team and present a taxonomy that we feel highlights many important open challenges in multi-agent research.
\end{abstract}


\end{frontmatter}

%%%%%%%%%%% The article body starts:

\section{Introduction}\label{sec:intro}

While multi-agent research pervades much of the work at DeepMind, largely due to work on solving grand challenges in games~\cite{SilverSSAHGHBLB17, VinyalsBCMDCCPE19}, the Game Theory \& Multi-Agent team, in particular, focuses on key aspects of multi-agent research. From our perspective, multi-agent research extends along several axes including
\begin{itemize}
    \item Number of Players: 2, >2, many, infinite
    \item Players: agents, agents + humans
    \item Payoffs / Incentives: competitive, mixed-motive, cooperative
    \item States: Discrete, Continuous
    \item Time: One-shot, Repeated / Iterated, Extensive-form, Infinite
    \item Observability: Perfect Information, Imperfect Information
    \item Inspiration: Game Theory, Evolution, Sociology.
\end{itemize}

Individual projects often work in settings or domains that vary along several of these dimensions at once. In the following sections, we will describe the areas of research prioritised within the group.
%
% (1) the technical problems in multi-agent systems tackled by the group (research agenda), including applications and industry collaboration;
% (2) the main approaches developed by the group and any key results achieved; and
% (3) a description of important open challenges in multi-agent systems research as viewed or prioritised by the group. (n-player, general-sum games, agent cooperation (with humans), mechanism design)
%
These include the computation of fundamental solution concepts from game theory such as equilibria in \S\ref{sec:eq}, the development of key multi-agent skills like negotiation in \S\ref{sec:skills}, the ability to shape the outcomes of multi-agent systems via mechanism design in \S\ref{sec:shape_outcomes}, novel, game-theoretic approaches to evaluating performance of multi-agent systems in \S\ref{sec:eval}, approaches inspired by nature and the success of human cultural evolution in \S\ref{sec:nature}, the formulation of fundamental problems in machine learning as games in \S\ref{sec:gamify}, the development and pursuit of solving challenging benchmarks to further push multi-agent research in \S\ref{sec:domains}, and multi-agent research aimed at solving current real world problems in cooperation in \S\ref{sec:good}.
% multi-agent research in challenging strategic, high-dimensional, and continuous state-action domains like humanoid soccer \S\ref{sec:control}, 

% \textcolor{red}{Do we need an additional dimension for mechanism design or can that be folded into asymmetry of agents?}

%\subsection{}\label{s1.1}

%%%%%%%%%%%%%%%%%%%%%%%%%%%%%%%%%%%%%%%%%%%%%%%%%%%%%%%%%%%%%%%%%%%%%%%%%%%%%
%%%%%%%%%%%%%%%%%%%%%%%%%%%%%%%%%%%%%%%%%%%%%%%%%%%%%%%%%%%%%%%%%%%%%%%%%%%%%

\section{Computing Equilibria / Solution Concepts}\label{sec:eq}
This line of work focuses primarily on selecting strategies based on outcomes in a game as opposed to strategies selected according to behavioural criteria (discussed next in \S\ref{sec:skills}). This includes the famous equilibrium concepts of Nash equilibrium (NE), correlated equilibrium (CE), and coarse-correlated equilibrium (CCE) but also others of our own design. We also differentiate between two-player, zero-sum games where these equilibrium concepts are well understood and $n$-player, general-sum games where they serve as a guide toward strategic play but do not necessarily provide a full-answer. Our group is particularly interested in exploring these equilibrium concepts and others in the $n$-player, general-sum setting. By studying these concepts empirically in a variety of games, we expect to generate insights that will extend our understanding of equilibrium concepts and performant strategies in $n$-player, general-sum games.

\subsection{Two Player, Zero-Sum}
In two-player, zero-sum games, Nash equilibrium is widely accepted as a suitable solution concept. While several algorithms offer convergence guarantees to a Nash equilibrium in simple game classes such as normal-form games, there are still opportunities for research in more complex classes such as imperfect information games.

% RL techniques for two-player zero-sum imperfect information games. Early methods were based on RL variants of classical fictitious play~\cite{Heinrich15FSP,Heinrich16NFSP}. Policy-Space Response Oracles (PSRO) generalised the idea to decouple the best response steps and meta-game solver~\cite{Lanctot17PSRO,muller2020generalised}.

One approach we take is to design an algorithm to maximise reward against a worst-case opponent~\cite{Lockhart19ED} or an improved opponent~\cite{Munos20} with final iterate convergence to Nash equilibria. Other methods draw inspiration from counterfactual regret minimization~\cite{Srinivasan18RPG,gruslys2020advantage} and replicator dynamics~\cite{hennes2020neural}, whose time average converges to Nash equilibria.

% , and reward transformations in regularised follow-the-leader~\cite{perolat2020poincare}.
In~\cite{perolat2020poincare}, we show that the popular naive learning approach of Follow the Regularized Leader (FoReL) cycles and fails to converge to the Nash equilibrium in sequential, imperfect information games. However, if we modify the players' payoffs by adding a regularization term that we anneal over iterations, we can recover convergence guarantees of the last iterate. We call this approach \emph{friction}-FoReL or \emph{F-FoReL} and demonstrate that this approach can be used to train state-of-the-art model-free reinforcement learning agents for two-player, zero-sum, sequential, imperfect information games.

% Stratego (perolat@, hennes@)[need to be careful cannot mention this yet, but maybe something on IIG]

\subsection{N-Player, General-Sum}\label{np_gensum}

In games with more than two players and in general-sum games, it is less clear what optimal play entails. There are several solution concepts that are still well defined in this setting and have been studied in various contexts. Our group works on computing these solution concepts and their variants as well as defining new ones.

\subsubsection{Nash Equilibria}
ADIDAS~\cite{gemp2021sample} stochastically approximates the Nash equilibrium of a normal-form game with many players, each with many actions. It does this by tracing a homotopy of quantal response equilibria defined with decaying levels of noise disturbance (equivalently, entropy bonus). For each entropy bonus level, the Nash equilibrium of the perturbed game is approximated by minimising a suitable energy function (a measure of $\epsilon$-Nash) with gradient descent. Critically, individual entries of the payoff tensor are only sampled as needed obviating the need to store the entire tensor in memory. This approach was used to approximate a unique Nash equilibrium (the limiting logit equilibrium) of a game with 7-players and 21-actions amounting to a payoff tensor with over 12 billion entries.

\subsubsection{(Coarse) Correlated Equilibria}
% (C)CE - (ian: this is potentially useful framing for a transition from NE to CE section, but note you don't have to argue a case for CE here. NE and CE are both established concepts in game theory).
% NE is a fundamental solution concept for two-player, constant-sum games. All NE have interchangeable solutions, equal and equal payoffs in this restricted setting. Furthermore, the solution can be found efficiently using linear programming. However, NE is not suitable for games with more than two players, or general-sum games.

% (marris@) Related solution concepts, correlated equilibrium (CE) and coarse correlated equilibrium (CCE), are more suitable for N-player, general-sum games. However, (C)CEs permit many valid solutions and selecting uniquely between them is known as the equilibrium selection problem.

Maximum Gini (Coarse) Correlated Equilibrium (MG(C)CE) \cite{marris2021_jpsro_icml} is a Tsallis-entropy-maximising equilibrium selection criterion. MG(C)CE uniquely selects an equilibrium from the convex polytope of valid (C)CE solutions and is invariant to positive affine transformations of the payoffs. It is easy to solve via existing off-the-shelf quadratic programming optimisers. MG(C)CE has a particularly compact dual representation if it selects a full-support solution.

% It is more computationally efficient that the related Maximum Entropy (Coarse) Correlated Equilibrium (ME(C)CE) \cite{ortix2007_mece}.

\subsubsection{Stationary Distributions of Fundamental Dynamics}

\emph{Replicator dynamics} is a key concept from evolutionary game theory that describes the dynamics of a population of strategies over time. In essence, they have been shown to describe a learning process and various reinforcement learning algorithms have been shown to converge in the limit to some version of a replicator dynamic \cite{Borgers97,TuylsVL03,BloembergenTHK15}. As such, replicator dynamics serve as a formal basis to understand the behaviour of various multi-agent learning algorithms and as a basis to develop new multi-agent learning algorithms \cite{TuylsHNM03,hennes2020neural}. We study the properties of the stationary distribution of these dynamics and how their equilibria might serve as a tractable solution concept in large games~\cite{omidshafiei2019alpha}.

% $\alpha$-Rank is an evolutionary dynamics based method that provides an alternative to the Nash equilibrium solution concept, allowing to capture dynamic behaviours in games and ranking of strategies.

The main advantages of this new concept concerns its uniqueness and efficient computation in many-player and general-sum games, making it a promising foundation for new multi-agent learning algorithms. The $\alpha$-regularised replicator dynamics define an irreducible Markov chain over a strategy set, which is called the \emph{response graph} of the game \cite{lanctot2017unified}, by indicating when a player has an incentive to make a unilateral deviation from their current strategy. The ordered masses of this Markov chain's unique stationary distribution yields the solution concept. The Markov transition matrix is directly linked to a solution concept called Markov-Conley chains (MCC)~\cite{omidshafiei2019alpha,PapadimitriouP18}. The correspondence of the stationary distribution to the MCC solution concept occurs in the limit of infinite $\alpha$.

\subsection{Infinite Number of Players}
As solving games with a high number of players becomes quickly intractable, an innovative approach for anonymous symmetric games considers instead the asymptotic limit where the number of players becomes infinite. This gives rise to the so-called Mean Field Games introduced in \cite{MFG_lasry-lions,2006huang-mfg} with numerous applications in crowd dynamics \cite{dogbe2010modeling}, finance \cite{carmonafouquesun2015mean}, epidemiology \cite{elie2020contact} or energy management \cite{hub2021mean} among others. Under structural conditions, a Nash policy of the Mean Field Game provides a good approximation for the solution of the corresponding $n$-player game, with an error vanishing as $n$ goes to infinity. For this reason, learning an equilibrium in Mean Field Games shall pave the way for scalable algorithms in multi-agent systems with a large population of players. 

We introduced and studied the convergence to a Nash equilibrium of several scalable learning algorithms for Mean Field Games, together with more specific applications to flocking \cite{perrin2021mean} or vehicles traffic routing management \cite{cabannes2021solving}. Our algorithms rely on Fictitious Play \cite{elie2020convergence,perrin2020fictitious} or Online Mirror Descent algorithms \cite{perolat2021scaling} and can be efficiently combined with Deep Reinforcement Learning \cite{lauriere2022scalable}. Using a well chosen enlarged class of policies, pinned as \emph{master policies}, allows us to generalise efficiently to several initial population distributions \cite{perrin2021generalization}. All known Nash-converging algorithms require some conditions on the game in order to work\textemdash monotonicity and contractivity being the most popular. Mean-Field PSRO~\cite{muller2022learning} provides a method to converge to a Mean Field Nash equilibrium, correlated and coarse-correlated equilibria in \emph{all games} through the use of accelerated adversarial regret minimization.
%We extended and scaled in applications such as flocking or vehicles traffic routing management. 
%\begin{itemize}
%    \item \textcolor{red}{Mean-field Games (relie@)}
%   
%    \item Mean-Field PSRO: Finding equilibria in Mean-Field Games is still an open subject. All known Nash-converging algorithms require some conditions on the game in order to work - monotonicity and contractivity being the most popular. Mean-Field PSRO~\cite{muller2022learning} provides a method to converge to Mean-Field Nash, correlated and coarse-correlated equilibria in \emph{all games} through the use of accelerated adversarial regret minimization.
%\end{itemize}

\subsection{Inspired Approaches for Very Large (Infinite) Strategy Spaces}

Much of our work into training agents to play games is inspired by iterative self-play approaches like Fictitious Play~\cite{brown1951iterative} and Double-Oracle (DO)~\cite{mcmahan2003planning}. We have extended these approaches into settings where strategies are represented by neural networks. In this case, the strategy space of each agent is the space of learnable parameters which is typically $\mathbb{R}^d$, and therefore infinite. Nevertheless, we have found that self-play approaches with approximate best responses perform well despite their lack of guarantees.

Early approaches that adapted to the multi-agent reinforcement learning setting were based on Fictious Play~\cite{Heinrich15FSP,Heinrich16NFSP}. A subsequent approach, policy-space response oracles (PSRO)~\cite{Lanctot17PSRO} extends Double-Oracle by approximately computing best responses with reinforcement learning. In more detail, PSRO constructs \emph{meta-games} where each action in the meta-game corresponds to a best response (``oracle'') policy, typically represented by a neural network. At each step, a Nash equilibrium of this meta-game is computed, after which a best response is approximated with (single-agent) deep reinforcement learning (RL) and then added to the meta-game strategy set.

$\alpha$-PSRO~\cite{muller2020generalized} alters the core algorithm of PSRO by replacing Nash equilibirum as the relevant meta-game solution concept with the stationary distribution of replicator dynamics developed in $\alpha$-Rank~\cite{omidshafiei2019alpha}. We show this change allows $\alpha$-PSRO to converge towards this particular stationary distribution, and demonstrate this empirically.

Similarly, (JPSRO) Joint PSRO~\cite{marris2021_jpsro_icml} replaces the Nash equilibrium with a (C)CE in an $n$-player, general-sum normal-form game. It benefits from the scaling properties of PSRO while converging to a (C)CE, a solution concept more suitable for cooperating agents.

One drawback of standard DO / PSRO approaches are that they scale poorly in real-world games that call for approximate best-response operators such as deep RL. Neural Population Learning (NeuPL, \cite{liu2022neupl}) makes progress towards bringing game-theoretic population learning algorithms to large games, leveraging the expressiveness of neural networks. Specifically, a single conditional network is used to explore, represent and optimise a ``live" population of strategies that are concurrently updated to best respond to mixed subsets of the population. By virtue of this shared representation, NeuPL enables positive transfer across strategies yet retains convergence guarantees to solution concepts such as NE. At convergence, NeuPL only retains a sequence of exact best-responses, yielding a compact representation of the strategically relevant policy space of the game.

%\textcolor{red}{Mean-field: (relie@) normalising flow work?} Not sure it really fits here or will be easy to present. 

Best-response policy iteration (BRPI) is a family of policy iteration methods~\cite{lagoudakis2003LSPI, SilverSSAHGHBLB17, anthony2021expert} that use an approximate best response calculator for policy improvement. Members of this family include different approximations to fictitious play. We applied BRPI to the board game Diplomacy~\cite{anthony2020learning}. This domain has a combinatorial action space, meaning that a player has to choose its own action from many possibilities (median number of possibilities in movement phases is $10^{45.8}$), and evaluate its performance against many more. We introduced sampled best response, an approximate best response calculation technique, to scale BRPI to this game.

%%%%%%%%%%%%%%%%%%%%%%%%%%%%%%%%%%%%%%%%%%%%%%%%%%%%%%%%%%%%%%%%%%%%%%%%%%%%%
%%%%%%%%%%%%%%%%%%%%%%%%%%%%%%%%%%%%%%%%%%%%%%%%%%%%%%%%%%%%%%%%%%%%%%%%%%%%%

\section{Developing Multi-Agent Skills}\label{sec:skills}
While solution concepts provide an axiomatic approach to selecting strategies in games, they currently do not provide a full satisfactory answer, as we mentioned, in the $n$-player, general-sum setting and in some cases, are intractable to compute. Given this shortcoming, several general strategies or skills have been identified as critical for navigating the complex interactions of multi-agent systems, for example, negotiation and team-formation. Developing these multi-agent skills provides agents with abilities to help them perform well in a variety of settings including interactions with humans.

\subsection{Team Formation and Alliances}
Cooperation between intelligent agents requires reasoning about what other agents might do, finding potential partners, forming teams and a joint plan of actions and negotiating how to share gains resulting from the collaboration. The game of Diplomacy~\cite{calhamer1959diplomacy} is a prominent AI challenge where identifying a mutually beneficial joint plan is crucial to winning the game. We have shown how reinforcement learning methods can deal with the large action spaces that occur in this game by applying a custom improvement operator that considers the actions that others might take~\cite{anthony2020learning}.

Beyond modeling the actions of other agents, organising into teams or coalitions provides structure that can facilitate collective action, even for a population of self-interested agents.   
Towards this end, we have examined direct mechanisms for negotiating team formation and showed that reinforcement learning can allow agents to construct coalitions and share the gains in a fair manner, similar to power indices from cooperative game theory~\cite{bachrach2020negotiating}.

%% Alliance Dilemmas
The process of bilateral alliance formation induces a social dilemma between the parties. Teaming up with peers requires an element of trust when peers might be incentivised to defect from a joint plan so as to increase their reward at the expense of others. We have found that independent multi-agent reinforcement learning algorithms fail to discover beneficial alliances in simple models of economic competition. When the agents are augmented with a peer-to-peer contract mechanism they learn to discover and enforce alliances \cite{hughes2020learning}.

%% \ Alliance Dilemmas

% \begin{itemize}
%     % \item Negotiating team formation (yorambac@) - Done
%     % \item \textcolor{red}{Alliance Dilemmas (edwardhughes@, yorambac@)} - Done (see above)
%     % \item Diplomacy (twa@, yorambac@) - Done...ish?
%     % \item Coalition Structure Generation (ksbullard@, katelarson@, yorambac@) - Hinted at it (difficult to go in depth since this work is in its infancy)
% \end{itemize}

\subsection{Evolution of Cooperation}
In order for multi-agent systems to survive in diverse and unknown environments, they must learn when and how to cooperate. In~\cite{mckee2020social}, we draw on interdependence theory from social psychology and imbue reinforcement learning agents with Social Value Orientation (SVO), a flexible formalization of preferences over group outcome distributions.
% We subsequently explore the effects of SVO on populations of reinforcement learning agents in two mixed-motive Markov games.
We demonstrate that heterogeneity in SVO generates meaningful and complex behavioural variation among agents, matching predictions made by interdependence theory. Empirical results in mixed-motive dilemmas suggest agents trained in heterogeneous populations develop particularly generalised, high-performing policies relative to those trained in homogeneous populations. Given the high fitness of diverse populations, it is possible evolutionary pressure selects for such heterogeneity.

In other work~\cite{gemp2022d3c}, we aim for agents to learn how to share rewards through interaction with their multi-agent environment. To derive such a learning rule, we first formulate a tractable measure of ``price of anarchy'', a technical, game-theoretic definition that quantifies group-level inefficiency\textemdash it compares the welfare that can be achieved through perfect coordination against that achieved by self-interested agents at a Nash equilibrium. In the process of constructing a method to minimise an upper bound on this measure, we recover a variant of the celebrated Win-Stay, Lose-Shift strategy from behavioural game theory, thereby establishing a connection between the global goal of maximum welfare and an established agent-centric learning rule. We demonstrate that this method improves outcomes for each agent and the group as a whole in several social dilemmas.
% including a traffic network exhibiting Braess's paradox, a prisoner's dilemma, and several multi-agent domains.

\subsection{Coordination}
A desirable property of our agents is that they are able to coordinate and collaborate well with novel partners. This is particularly relevant to robotics and AI assistants which are meant to work with, or augment, humans in real-world tasks.
Zero-shot coordination (ZSC) examines the problem of enabling agents to generalise their coordination strategy to held-out partners, for example other independently trained agents or humans.

The game of Overcooked simulates a collaborative cooking environment and has recently been proposed as a coordination challenge for AI \cite{carroll2019utility}.  For this domain, we introduced Fictitious Co-Play (FCP) \cite{strouse2021collaborating}, a state-of-the-art method which trains an agent to collaborate with a diverse set of training partners. FCP was able to show strong performance when paired with both novel AI agents and human partners. Notably, our method was able to achieve this performance without the expensive process of collecting human data for the training process. This is a promising result towards efficiently producing agents which are capable of assisting and cooperating with humans in novel contexts.

% \begin{itemize}
    % \item \textcolor{red}{Zero-shot coordination: Overcooked (reverett@, ksbullard@)} - Done
% \end{itemize}

\subsection{Agents that Model Social Opponents and Norms}
In order to design agents capable of interacting with humans in society, it is imperative that they be able to learn how to interpret and adapt to human behaviour.

In zero-sum games, there are often intransitivities in strategy space \cite{czarnecki2020real} (e.g., rock beats scissors beats paper beats rock). Modelling and understanding other players' incentives becomes critical to appropriately and quickly adapt to unknown opponents. We have developed techniques inspired by hierarchical reinforcement learning to train agents that can infer other agents' policies and best respond to them as zero-shot generalisation \cite{vezhnevets2020options,kopparapu2022hidden,moreno21}. These techniques are a form of centralised training and decentralised execution, where privileged information is only available in hindsight to aid the training of representations.

Beyond modeling of individual motivations, agents embedded in a social context need to understand and appropriately respond to society-wide phenomena like norms. Our research has developed new algorithms to imbue our agents with the ability to adapt to norms~\cite{vinitsky2021learning}, as well as using these agents as models of human behaviour to understand the emergence of norms and institutions (including those that appear arbitrary or maladaptive) from first principles \cite{koster2022spurious}.

\subsection{Imitating other Agents}\label{subsec:imitation}
Human intelligence is especially dependent on our ability to acquire knowledge efficiently from other humans. This knowledge is collectively known as culture, and the transfer of knowledge from one individual to another is cultural transmission. Cultural transmission leads to a feedback loop bootstrapping individual and collective abilities, which in 12 millennia has taken us from hunting and gathering to international video calling. This suggests we endow agents with the ability to transmit their knowledge across the multi-agent system, so that subsequent ``generations" may grow in intelligence.

Inspired by this, we have used deep reinforcement learning to generate artificial agents capable of test-time cultural transmission \cite{team2022learning}, particularly focusing on real-time third-person imitation of other agents. Our trained agent can infer and recall navigational knowledge demonstrated by experts, purely via a real-time experience stream from a multi-agent RL environment. This knowledge transfer generalises across a vast space of previously unseen tasks. For example, our agents quickly learn new behaviours by observing a single human demonstration, without ever training on human data. This paves the way for cultural evolution as an algorithm for developing more generally intelligent artificial agents.

\subsection{Language and Communication}\label{sec:language}

Language has been mostly explored through a data-driven paradigm over the last decade~\citep{rae2021scaling}. However, language, and more broadly communication, is intrinsically a multi-agent problem: communication requires at least two interacting agents per se. In this spirit, our team focused on two facets of language in multi-agent settings: language emergence and improving language processing methods.

Language emergence explores how a communication protocol may emerge and evolve when multiple agents must cooperate to solve a task. From a scientific perspective, such computational approaches aim to simulate the prerequisite or processes that could trigger the emergence of a structured language and shed light on human language evolution. From a machine learning perspective, it provides a novel benchmark to analyze agent dynamics and human-machine interactions. In this spirit, we explore how a large population may entail different language properties, either emphasising the importance of population heterogeneity~\cite{rita2021role} or population dynamics in large-scale experiments~\cite{chaabouni2021emergent}. We also considered the impact of embodied communication, i.e., when communication occurs within small interactive worlds. There, we highlight how simple human biases help promote communication in cooperative  tasks~\cite{eccles2019biases,kalinowska2022situated}.

On the other hand, game-theoretic solutions can also be used to build better machine learning models for natural language processing systems. For instance, we show that vocabulary selection can be cast as a team formation game between the possible words. Power indices from, e.g., the Shapley value and Banzhaf index cooperative game, allow finding small vocabularies, significantly reducing memory and compute resources while retaining high performance. Concurrently, we show that using strategies to dynamically prune vocabulary allow training a language model from scratch by solely relying on interaction~\cite{donati2021learning}.

%%%%%%%%%%%%%%%%%%%%%%%%%%%%%%%%%%%%%%%%%%%%%%%%%%%%%%%%%%%%%%%%%%%%%%%%%%%%%
%%%%%%%%%%%%%%%%%%%%%%%%%%%%%%%%%%%%%%%%%%%%%%%%%%%%%%%%%%%%%%%%%%%%%%%%%%%%%

\section{Shaping Outcomes of Multi-Agent Systems}\label{sec:shape_outcomes}
% This line of work looks at games where there is a strong asymmetry between players payoffs and / or strategy spaces. In particular, these works look at auctions and settings where a single coordinator seeks to shepherd the remaining players toward a desired solution.

In some games, there may be a strong asymmetry between player payoffs / strategy spaces where some players have privileged action over others. This includes settings such as auction design where a player called the \emph{designer} specifies the rules of an auction in order to elicit certain behaviour from the remaining players called \emph{participants}. Seminal results in auction design typically study settings with finite numbers of discrete outcomes. In our group, given DeepMind's work on deep reinforcement learning, we studying mechanism design in the multi-agent reinforcement learning setting where agents might have preferences over the weight space of neural networks (i.e., items are points in $\mathbb{R}^d$). The amount of asymmetry between players varies over domains we study.

\subsection{Static-Mechanisms}
In some settings, we learn a static mechanism once and deploy it. For instance, we have shown how neural networks can be used as a language for expressing agent preferences where these preferences are given over continuous spaces~\cite{bachrachneural}. This enables applying frameworks from the field of mechanism design, such as the VCG auction~\cite{milgrom2004putting} to reach decisions that maximise the welfare of all the agents.
    
Auctions are the protocol of choice to allocate goods to strategic buyers that have preferences over them. While researchers have proposed reliable auction rules that work in extremely general settings, these protocols may require extracting high payments from participants so as to ensure incentive constraints are maintained. This may not always be desirable, particularly in situations where the "auctioneer" is only interested in the efficient allocation of resources and does not care about revenue. By casting auction rule design as a statistical learning problem, we trade generality for participant welfare effectively and automatically, using a novel deep learning network architecture and auction representation \cite{tacchetti22learning}. Our analysis shows that the resulting auction rules outperform state-of-the art approaches in terms of participants' welfare, applicability, robustness.

We have also analysed how reinforcement learning agents can be employed as the ``inner loop'' in an empirical approach to mechanism design. Many real-world tasks have spatial and temporal intricacies that cannot readily be reduced to matrix games. In \cite{bakker21}, we examined how interventions on geometry, cost, specialization and topology affected the emergence of reciprocal care in a spatialised, temporally-extended network game. The effects of mechanism choice on agent learning were subtle and sometimes counter-intuitive, pointing at a need for more detailed modelling of mechanistic interventions, especially in an increasingly socio-technical world.

\subsection{Dynamic-Mechanisms}
In other work, we deploy a mechanism online which must learn from participants' behaviour and adjust itself on the fly in order to elicit the desired outcomes. This setting brings several additional challenges including complex effects of hysteresis where the rules of the mechanism early in the lifetime of agents affects their behaviour later in their lifetime.

Evaluation in multi-agent settings focuses primarily on the interaction among fixed, non-learning co-players. While this evaluation methodology has merit, it fails to capture the dynamics faced by agents interacting with continually learning co-players. The Good Shepherd~\cite{balaguer22shepherd} addresses this limitation, and constructs agents (``mechanisms'') that perform well when evaluated over the learning trajectory of their adaptive co-players (``participants''). The algorithm consists of two nested learning loops: an inner loop where participants learn to best respond to fixed mechanisms; and an outer loop where the mechanism agent updates its policy based on experience.

HCMD-{\it zero}~\cite{balaguer22hcmd} is a general-purpose method to construct mechanism agents that are preferred over baseline alternatives by human participants engaged in economic interactions. HCMD-{\it zero} learns by mediating interactions among participants, while remaining engaged in an electoral contest with copies of itself, thereby accessing direct feedback from participants. Results on the Public Investment Game, a stylised resource allocation game that highlights the tension between productivity, equality and the temptation to free-ride, show that HCMD-{\it zero} produces competitive mechanism agents that are consistently preferred by human participants over baseline alternatives, and does so automatically, without requiring human knowledge, and by using human data sparingly and effectively.

We have also considered the setting where a single agent might learn to reward others, thus eliciting desirable cooperative behaviours. In the Learning to Incentivize Others (LIO) algorithm \cite{yang2020learning}, the agent has a ``gifting'' policy represented as a neural network. The parameters of this network are adjusted to maximise the original environment reward (without gifts) and regularised to approximately maintain budget-balance.
% LIO is capable of maximising welfare via division of labor on a version of the Cleanup game~\citep{hughes2018inequity}, a standard environment used to model a public goods dilemma in a gridworld setting.

%%%%%%%%%%%%%%%%%%%%%%%%%%%%%%%%%%%%%%%%%%%%%%%%%%%%%%%%%%%%%%%%%%%%%%%%%%%%%
%%%%%%%%%%%%%%%%%%%%%%%%%%%%%%%%%%%%%%%%%%%%%%%%%%%%%%%%%%%%%%%%%%%%%%%%%%%%%

\section{Evaluating / Analysing Players \& Strategies}\label{sec:eval}
Evaluating the performance of players or strategies in a game is not always straightforward. In a sport like bobsledding, teams' performances are largely independent of each other and so teams can be ranked by finishing time. However, in a game like football where two teams play head-to-head in various stadium environments, performances are intertwined. This is made evident in the classic rock-paper-scissors game where each strategy wins, ties, and loses to exactly one other strategy. Given a game with arbitrary payoffs, how can we score or rank the strategies available to us in that game? We are especially interested in game-theoretic approaches that satisfy certain desirable properties. Along similar lines, we aim to use principles from game theory to statistically evaluate the performance and compatibility of players in real world sports settings.

\subsection{Rating}

Rating strategies in normal-form games is an important aspect of evaluating agents in multiplayer games and tasks in multitask scenarios. Normal-form games can be constructed empirically from measured performance between matchups of agent policies, where each policy corresponds to a player's strategy in the game. Such empirical games are often called \emph{meta-games}, and are utilised in empirical game-theoretic analysis (EGTA) and PSRO. Traditionally, only transitive dependencies between strategies were considered when defining a rating (e.g. Elo), while cyclic (``rock-paper-scissors'') dynamics were ignored. Recent work~\cite{balduzzi2018re} provided a mechanism to effectively rate strategies in two-player zero-sum games. The approach taken there was to compute the Nash equilibrium of this metagame and then rank strategies either according to their mass under the equilibrium or their \emph{Nash average}. Our work in~\cite{marris2021_jpsro_icml} generalises Nash averages to player payoff gradients at the solution (e.g., Nash equilibrium) of the game. Each of the solution concepts described in \S\ref{np_gensum} can be used to define a new rating or ranking procedure in a similar manner. For example, $\alpha$-Rank uses the ordered masses of replicator dynamic's unique stationary distribution to provide strategy profile rankings.

\subsection{Sports Analytics}

The availability of team sports data (e.g., in football or basketball) has been steadily increasing over recent years. Football, in particular, is a great test bed to study novel AI techniques with the potential to offer decision-makers advice in the form of an automated video-assistant coach (AVAC)~\citep{tuyls2021game}. Techniques from computer vision, statistical learning, and game theory all contribute to the understanding of the game. Players can be tracked in broadcast video using computer vision; event and pose detection techniques can provide further context. Statistical learning allows constructing player representations that capture individual playing style and intra-team chemistry, and also allows estimating the contribution of individual player actions. Game theory provides the framework and tools to study the sequential decision-making problems players face in the presence of other players (cooperative and adversarial).

The AVAC system sits at the intersection of these three research fields. In~\citep{tuyls2021game}, we identify three frontiers: video predictive models, strategic video generation models, and interactive decision-making, and lay out a roadmap of specific research and engineering problems. The development of AI-powered sports analytics techniques offer great potential for decision-makers, broadcasters, players, and fans.

\section{Inspired by Nature, Biology, and Sociology}\label{sec:nature}
Many of the most outstanding forms of intelligence that we see in the real world have arisen from multi-agent systems. For example, cultural evolution is thought to have been crucial to social intelligence and our ability to cooperate in large numbers and solve complex problems (e.g., via market economies). For this reason, some of our research looks to replicate the forces in nature that are believed to have led to humanity's and nature's exponential development. This is particularly evident in the work we described in \S\ref{subsec:imitation}.

Living systems show a striking ability to adapt to novel situations and to produce ever more complex and sophisticated forms of life. It is only through (social) interactions that individuals have acquired these abilities, be they implicitly through a process of natural selection, or explicitly in the form of learning \citep{szathmary1995major,ginsburg2021evolutionary}. Social interactions can lead to autocurricula (that is, a curriculum that automatically matches the skill level of the individual) because the task to learn or adapt to is precisely the interaction with others \cite{leibo2019autocurricula}. We use nature as inspiration for designing learning systems that are able to leverage social interactions in their learning path either implicitly through pure interactions \cite{LeiboZLMG17, perolat2017multi}, or explicitly via an evolutionary or ecological level above reinforcement learning \cite{wang2018evolving,leibo2019malthusian,sunehag2019reinforcement}. We also model known or putative human inductive biases to achieve more human-like behaviours \cite{hughes2018inequity,jaques2019social,mckee2020social}, which result in agents that can escape defecting equilibria in social dilemmas.

%%%%%%%%%%%%%%%%%%%%%%%%%%%%%%%%%%%%%%%%%%%%%%%%%%%%%%%%%%%%%%%%%%%%%%%%%%%%%
%%%%%%%%%%%%%%%%%%%%%%%%%%%%%%%%%%%%%%%%%%%%%%%%%%%%%%%%%%%%%%%%%%%%%%%%%%%%%

\section{Gamification}\label{sec:gamify}
The bulk of modern AI and machine learning is built on the foundation of learning as optimization. Given how pervasive multi-agent intelligence is in the natural world, we suspect a form of learning intrinsically tied to game theory and multi-agent research is warranted. Our team explores several lines of work that attempt to reformulate fundamental aspects of learning as games.
% and communication 

We aim to reformulate fundamental problems in machine learning as solving games. For example, in~\cite{gemp2020eigengame, gemp2021eigengame}, we reformulate the problem of singular value decomposition (SVD) as finding the Nash equilibrium of an \emph{EigenGame}, bearing similarity to a congestion game. In this case, players control eigenvectors in order to maximise the variance captured in a dataset minus a cost for their eigenvector aligning with other players'. SVD underlies several machine learning problems including principal component analysis, linear discriminant analysis, latent semantic indexing, spectral clustering, proto-value functions, and others. Formulating SVD as a game makes it obvious how to design an algorithm that distributes over players.
% EigenGame is currently state-of-the-art in the literature for approximating SVD for very large datasets (e.g., $\approx 100$ TB).

We also revisit reinforcement learning problems where the agent's utility is concave in the occupancy measure induced by its policy (CURL) and formulate them as Mean Field Games \cite{geist2021curl}. CURL encompasses frameworks such as (pure) exploration, imitation learning, marginal matching or constrained MDPs, that are efficiently solved using algorithms recently introduced for solving Mean Field Games \cite{elie2020convergence,perolat2021scaling,muller2022learning}. This also provides new convergence rates by drawing connections between Fictitious Play and Frank-Wolfe. %Last sentence can be removed if shortening is necessary.

Lastly, in \S\ref{sec:language} we mentioned how we used Shapley values from cooperative game theory to optimally select features, in this case vocabularies, in language tasks~\cite{patel2021game}.

% Argues language / communication is inherently multi-agent.
%We have considered how agents might identify a ``language'' (joint code) in cooperative tasks, finding biases that help shape reinforcement learning algorithms so as to allow agents to effectively communicate to solve a joint task~\cite{eccles2019biases}, showing standard reinforcement learning methods without such biases may fail to deal even with relatively simple tasks that require communication.

%\textcolor{red}{Language and games (fstrub@)} Scaling such \emph{emergent communication} approaches to large populations is particularly difficult, but important for understanding human language evolution and building more efficient languages. We focus on three independent aspects to scale up, namely the dataset, task complexity, and population size~\cite{chaabouni2021emergent}.

% \subsubsection{Open-ended Learning / Environments as Agents and Vice Versa}
% \begin{itemize}
%     % Ada is not public knowledge -- \item Ada (edwardhughes@, vibhadasagi@, avishkar@)
%     \item Open Ended Learning + Interaction Graph (garnelo@)
% \end{itemize}

%%%%%%%%%%%%%%%%%%%%%%%%%%%%%%%%%%%%%%%%%%%%%%%%%%%%%%%%%%%%%%%%%%%%%%%%%%%%%
%%%%%%%%%%%%%%%%%%%%%%%%%%%%%%%%%%%%%%%%%%%%%%%%%%%%%%%%%%%%%%%%%%%%%%%%%%%%%

\section{Test Suites and Benchmarks}\label{sec:domains}
% This line of work aims to develop games for testing the development of our multi-agent algorithms.
% In order to advance the state of multi-agent research, we think it is beneficial to have a diverse set of challenging benchmarks to test against. Each of these explores various dimensions of the multi-agent space as outlined in the introduction and we have open-sourced several of these in order to aid researchers around the world.

Testing benchmarks are key to the success of artificially intelligent systems. Rapid progress tends to occur whenever there is a public, objective way to measure the quality of a proposed solution. A prime example is the development of ImageNet \cite{deng2009imagenet} which together with AlexNet \cite{krizhevsky2012imagenet} launched the Deep Learning revolution. Reinforcement learning has comparatively fewer examples of this, and even fewer so in multi-agent reinforcement learning. Each of the projects below explores various dimensions of the multi-agent space as outlined in the introduction and we have open-sourced several of these in order to aid researchers around the world.

Melting Pot \cite{leibo2021scalable} is a test suite for evaluating zero-shot generalisation to novel social circumstances. We will continue to expand and improve the suite, intending to increase coverage of the breadth of social situations like: zero-sum games, general-sum games, games where the incentives of players are mostly aligned, and those in which they are mostly opposed, games of coordination, games of team competition, and many more.

Diplomacy is a decades old AI challenge designed to test game playing agents in a simultaneous move, extensive-form game setting with an incredibly large action space. The game is designed to emphasise the interactions between agents, meaning there is no universally effective strategy, and instead agents must choose a strategy compatible with those of their co-players. We specifically look at the no-press version of this game (no communication between agents). This game continues to serve as a benchmark for AI approaches today~\cite{gray2020human, bakhtin2021no, paquette2019no, anthony2020learning}.

Soccer is a long-standing grand challenge for AI because it captures many key open problems that must be solved on the route to useful physically embodied AI systems. Those challenges include: high-frequency, high-dimensional, continuous full body control; agile locomotion; reusing multiple skills in different contexts (simultaneously or seamlessly chained together); long-term planning and strategy; cooperating with others; being robust to a range of other agents' behaviours and adversarial situations. In the MuJoCo soccer project, we used soccer to demonstrate that a learning system which controls a simulated humanoid embodiment using 56-dimensional instantaneous joint torques, can learn mid-level skills (locomotion, getting up, kicking, dribbling) and ultimately some teamwork (division of labor, passing) in the service of long-term goals. Teams of agents were trained to play soccer via reinforcement learning and imitation of basic human motor control. They exhibit human-like motor control and complex behaviour at different scales, quantified by a range of analysis and statistics, including those used in real-world sports analytics to quantify teamwork~\cite{liu2021motor,liu2019boxhead}.

The reformulation of machine learning problems as games represents an opportunity for new domains of study. For example, EigenGame~\cite{gemp2020eigengame,gemp2021eigengame} is an $n$-player, general-sum game where each player's strategy space is constrained to the unit sphere, a Riemannian manifold. Because the strategy spaces are not finite as in a finite normal-form game, EigengGame does not enjoy their traditional properties. For instance, there is no symmetric Nash equilibrium in an EigenGame despite the fact that EigenGame is a symmetric game. The sorts of games that capture machine learning and other foundational mathematical problems may present new areas of study for game theorists and machine learning practitioners alike.

Language emergence simulation only recently started to scale up to medium-size populations with non-artificial inputs. This scaling-up paves the way to new agent dynamics for communication and raises novel multi-agent questions. What makes a language stable, evolve, or degenerate? Can we link language structure to game equilibria? We hence release a large-scale framework on language emergence to help the community towards this goal~\cite{chaabouni2021emergent}.

OpenSpiel~\cite{LanctotEtAl2019OpenSpiel} is an open-source framework we have released for reinforcement learning and search in extensive-form games. It supports $n$-player games, simultaneous move games, imperfect and perfect information, and mean-field games. It contains many basic implementations of algorithms listed in this document, including some visualizations and evaluation tools such as $\alpha$-Rank~\cite{omidshafiei2019alpha}.

%%%%%%%%%%%%%%%%%%%%%%%%%%%%%%%%%%%%%%%%%%%%%%%%%%%%%%%%%%%%%%%%%%%%%%%%%%%%%
%%%%%%%%%%%%%%%%%%%%%%%%%%%%%%%%%%%%%%%%%%%%%%%%%%%%%%%%%%%%%%%%%%%%%%%%%%%%%

\section{MA for Good}\label{sec:good}
% Our team also seeks to use multi-agent research to solve current real world challenges.
Much of our research into multi-agent systems can be tied to providing solutions to real-world problems, however, some have more direct applications. This line of work aims to concretely impact humanity in a positive way.

% YB: Ian, consider moving on earlier in this paper as a broad motivation.
A key part of our motivation from multi-agent research relates to aspiring to solve problems of cooperation between AI agents and their peers (both other AI agents or humans). As AI technology becomes more ubiquitous, it has the potential to improve collective behaviour in many domains, such as transport, consumer and financial markets, communication and social media, or security domains~\cite{dafoe2021cooperative}. Better understanding of multi-agent cooperation could yield benefits in both small-scale problems such as better traffic flows, and large-scale problems such as pandemic preparedness or global trading. We have surveyed key problems and challenges to unlock stronger ``Cooperative AI''~\cite{dafoe2020open}, including aspects of multi-agent {\it understanding} (such as predicting the goals and behaviours of others), {\it communication} and sharing of information, intents and preferences, {\it commitment} and making credible promises regarding agent behaviour and {\it norms and institutions}, such as shared beliefs or rules, which serve as social infrastructure that reinforces understanding, communication and commitment. The Cooperative AI Foundation (CAIF) has been set up as an independent body aimed at advancing this research agenda across disciplines.

Multi-agent reinforcement learning is a powerful tool for modelling human societies. This is particularly critical as many of the biggest challenges of our time are social in nature, like the emergence of discrimination. Statistical discrimination refers to the behaviour of estimating the quality of a social partner based purely on the statistical association between some readily perceptible characteristics and quality, without taking into account the specific quality of the individual in question \cite{arrow1972some}. In recent work \cite{duenez2021statistical}, we studied statistical discrimination in a temporally and spatially extended context where individuals have the chance to progressively refine their estimate of the quality of partners, and where the limitations of learning agents are explicitly modelled. We discovered that an agent's information processing ability mediates whether they will learn to individuate their partners and choose them based on their intrinsic quality, or fail to do so and engage in statistical discrimination.

\section{Conclusion}\label{sec:concl}
In this summary, we have briefly outlined the research pursued by the Game Theory \& Multi-Agent team at DeepMind. Our agenda explores several axes of the multi-agent problem space as described in \S\ref{sec:intro}. Each axis brings key technical challenges which we encounter in a variety of externally and internally defined domains. Given the expertise of DeepMind, much of our research takes a deep (neural network) multi-agent reinforcement learning approach, however, we also work on fundamental aspects of game theory such as computation of equilibria. Each of the sections in this summary represents important open challenges that our group has prioritised in order to efficiently push multi-agent research in a way that maximises impact on our understanding of multi-agent problems and their real world challenges.

%\begin{figure}[t]
%\caption{Figure caption.}\label{f1}
%\end{figure}

%\begin{table*}
%\caption{} \label{t1}
%\begin{tabular}{lll}
%\hline
%&&\\
%&&\\
%\hline
%\end{tabular}
%\end{table*}

%%%%%%%%%%% The bibliography starts:

%%  ios1.bst will be used to                               %%

% \nocite{*}
% if your bibliography is in bibtex format, use those commands:
\bibliographystyle{ios1}           % Style BST file.
\bibliography{bibliography}        % Bibliography file (usually '*.bib')

% or include bibliography directly:
%\begin{thebibliography}{0}
%\bibitem{r1} F. Author, Information about cited object.
%
%\bibitem{r2} S. Author and T. Author, Information about cited object.
%\end{thebibliography}

\end{document}